\documentclass[%
 reprint,
superscriptaddress,
groupedaddress,
 amsmath,amssymb,
 aps,
prb,
]{revtex4-1}

\usepackage{graphicx}
\usepackage{dcolumn}
\usepackage{bm}
\usepackage[colorlinks=false,pdfborder={0 0 .1}]{hyperref}

\begin{document}


\title{Coherent coupling between ferromagnetic magnon and superconducting qubit}

\author{Y. Tabuchi$^1$}
\author{S. Ishino$^1$}
\author{A. Noguchi$^1$}
\author{T. Ishikawa$^1$}
\author{R. Yamazaki$^1$}
\author{K. Usami$^1$}
\author{Y. Nakamura$^{1,2}$}

\affiliation{%
$^1$Research Center for Advanced Science and Technology (RCAST),\\
The University of Tokyo, Meguro-ku, Tokyo 153-8904, Japan
}%

\affiliation{%
$^2$Center for Emergnent Matter Science (CEMS), RIKEN, 
 Wako, Saitama 351-0198, Japan 
}%

\date{\today}


\maketitle

{\bf 
Rigidity of an ordered phase in condensed matter results in collective excitation modes spatially extending in macroscopic dimensions\cite{bib:Goldstone62}. 
Magnon is a quantum of an elementary excitation in the ordered spin system, such as ferromagnet.
Being low dissipative, dynamics of magnons in ferromagnetic insulators has been extensively studied and widely applied for decades in the contexts of ferromagnetic resonance\cite{bib:Sparks64,bib:Gurevich10}, and more recently of Bose-Einstein condensation\cite{bib:Demokritov06} as well as spintronics\cite{bib:Uchida08,bib:Kajiwara10}.
Moreover, towards hybrid systems for quantum memories and transducers, coupling of magnons and microwave photons in a resonator have been investigated\cite{bib:Huebl13,bib:Tabuchi14,bib:Zhang14,bib:Goryachev14}.
However, quantum-state manipulation at the single-magnon level has remained elusive because of the lack of anharmonic element in the system.  
Here we demonstrate coherent coupling between  a magnon excitation in a millimetre-sized ferromagnetic sphere and a superconducting qubit, where the interaction is mediated by the virtual photon excitation in a microwave cavity.
We obtain the coupling strength far exceeding the damping rates, thus bringing the hybrid system into the strong coupling regime. 
Furthermore, 
we find a tunable magnon-qubit coupling scheme utilising a parametric drive with a microwave.
Our approach provides a versatile tool for quantum control and measurement of the magnon excitations and thus opens a new discipline of quantum magnonics.
}

%
%
%
%
%

Single electron spins, being a natural and genuine two-level system, play crucial roles in numerous applications in quantum information processing. The intrinsic drawbacks, however, are its small magnetic moment $\mu_{\textrm{B}}$, the Bohr magneton, and the limited spatial extension of the electron wavefunction, making coherent coupling with an electromagnetic field rather weak. To circumvent the problems, paramagnetic spin ensembles have been actively studied using atoms\cite{bib:Hammerer10}, NV centres\cite{bib:Zhu11,bib:Kubo11}, and rare-earth ions in a crystal\cite{bib:Longdell05,bib:Probst13}. The coupling strength is largely enhanced by the square-root of the number of spins involved. At the same time, a collective spin excitation mode, which matches the input electromagnetic-field mode, is spanned in the spatially and spectrally extended ensemble. However, with an increased spin density for stronger coupling, the spin-spin interactions among the ensemble drastically degrade the coherence of the system and thus make a trade-off. 

We move one-step further by introducing ferromagnets. Even though they typically have a spin density several orders of magnitude higher, the strong exchange and dipolar interactions among the spins dominate their dynamics and form narrow-linewidth magnetostatic modes. The simplest mode has the uniform spin precessions of the rigid spins in the whole volume, called the Kittel mode. Coherent coupling between the Kittel-mode magnons and microwave photons in a cavity was recently demonstrated in the quantum regime\cite{bib:Tabuchi14}.

Superconducting qubits are also an excellent example of quantized collective excitations in macroscopic-scale electrical circuits, where the nonlinearity of Josephson junctions plays a crucial role for the realisation of the qubit, i.e., an effective two-level system. The progress in the last decade has made the qubits and their integrated circuits one of the most advanced technologies for quantum information processing\cite{bib:Devoret13,bib:Barends14,bib:D-wave}. In the setups of circuit quantum electrodynamics, a qubit as an artificial atom is coupled strongly to a microwave resonator\cite{bib:Wallraff04} or a waveguide\cite{bib:Astafiev10}. They allow precise control and readout of the qubit states as well as synthesis and characterisation of arbitrary quantum states in the microwave modes coupled to the qubits\cite{bib:Hofheinz09}. These techniques can readily be applied to quantum engineering of other physical systems.

In this Letter, we demonstrate a hybrid quantum system which combines two heterogeneous collective-excitation modes, i.e., the Kittel mode in a ferromagnetic crystal and a superconducting qubit.  The nonlinearity brought by the qubit is the critical element to allow non-classical control of magnon excitations.

Our experimental setup is illustrated in Fig.~\hyperlink{fig:fig1h}{1}. A transmon-type superconducting qubit and a single-crystalline yttrium-iron-garnet (YIG) sphere are mounted in a microwave cavity. The qubit with a 0.7-mm-long dipole antenna has a resonant frequency $\omega_{\textrm{q}}/2\pi$ of 8.136~GHz. It strongly couples to the electric fields of the cavity modes; e.g., the coupling strength $g_{\textrm{q}}/2\pi$ between the qubit and the TE${}_{102}$ (transverse electric) mode at $\omega_{102}/2\pi = 8.488$~GHz is 121~MHz (See \hyperlink{sec:A_qubit_and_cavity}{Supplementary Information} for other detailed parameters). 
The YIG sphere with a diameter of 0.5~mm is glued to an aluminium-oxide rod and mounted near the anti-node of the magnetic field of the TE${}_{102}$ mode. We also apply a local static field $B_{\textrm{static}}$ $\sim$ 0.29~T which makes the YIG sphere a single-domain ferromagnet~(Ext.~Fig.~\hyperlink{fig:exfig1}{1}). The sphere now has an enormous magnetic dipole moment $N\mu_B$ which couples strongly to the magnetic field of the cavity mode. The large enhancement factor $N$ = $1.4 \times 10^{18}$ is the number of net electron spins in the sphere; we take an advantage of the high spin density as compared to that of previously studied paramagnetic systems.

We perform a series of spectroscopic measurements in a dilution refrigerator at $T$ = 10~mK.  All the data are taken in the quantum regime where very few thermally-excited photons and magnons exist. The average probe-photon number in the cavity is also kept below one. 

To characterise the coupling between the magnon and photon, we perform spectroscopy with the qubit frequency for detuned. Figure~\hyperlink{fig:fig1h}{1b} shows the normal-mode splitting between TE${}_{102}$ mode and the Kittel mode in the YIG sphere. The pronounced anticrossing indicates the strong coupling between the two systems\cite{bib:Tabuchi14}. We obtain the coupling strength, $g_{\textrm{m}}/2\pi$=21.0~MHz, and the linewidths of the TE${}_{102}$ and Kittel modes, $\kappa_{102}/2\pi$=2.5~MHz and $\gamma_{\textrm{m}}/2\pi$=1.4~MHz, from the fit. The additional splitting seen in the upper branch originates from another magnetostatic mode which is detuned from the Kittel mode by 4.3~MHz and is coupled to the cavity mode with a strength of 4.2~MHz. 

While the qubit and the magnon electrically and magnetically couple to the cavity mode respectively, they have negligibly small direct interaction. 
Therefore, we first establish a static coupling scheme between the magnon and the qubit by utilizing the presence of the cavity mode (Fig.~\hyperlink{fig:fig2h}{2a})\cite{bib:Imamoglu09}. We tune the qubit and the magnon frequencies, $\omega_{\textrm{q}}$ and $\omega_{\textrm{FMR}}$, while both are far detuned from the cavity frequency $\omega_{\textrm{c}} (\equiv \omega_{102}$). When $\omega_{\textrm{q}} \simeq \omega_{\textrm{FMR}}$, coherent exchange of the qubit excitation and a magnon is mediated by the virtual-photon excitation in the cavity mode. The interaction is described by a Jaynes-Cummings-type Hamiltonian, which is written as
\begin{equation}
  {\cal \hat{H}}_{\textrm{qm,s}}/\hbar
        = g_{\textrm{qm,s}}\, \hat{\sigma}^{+} \hat{c}
        + g_{\textrm{qm,s}}^{*}\, \hat{\sigma}^{-} \hat{c}^\dagger,
  \label{eq:qubit-magnon}
\end{equation}
where $g_{\textrm{qm,s}} = g_{\textrm{q}}\,g_{\textrm{m}} / \Delta$ is the effective qubit-magnon coupling strength, and $\hat{\sigma^{-}}[=(\hat{\sigma}^{+})^\dagger]$ and $\hat{c}$ are annihilation operators of the qubit and the magnon, respectively. The detuning $\Delta$ is the difference between the bare frequencies of the qubit and the cavity mode (See \hyperlink{sec:A_static}{Supplementary Information}). The first and the second excited states of the hybridised system are bonding and antibonding states between the qubit and the magnon excitation. 

To demonstrate the qubit-magnon coupling, we perform qubit excitation spectroscopy by using the qubit readout through the cavity TE$_{103}$ mode. Despite the large detuning, the TE$_{103}$ mode at $\omega_{103}/2\pi = 10.461$~GHz is subject to a dispersive frequency shift: the change in the cavity reflection coefficient, Re($\Delta r$), at $\omega_{103}$ reflects the qubit state. Figure~\hyperlink{fig:fig2h}{2b} shows Re($\Delta r$) as a function of the excitation microwave frequency and the static magnetic field. The anticrossing is a manifestation of the magnon-vacuum-induced Rabi splitting, which indicates coherent coupling between the qubit and the Kittel mode. From the size of the splitting, the effective coupling strength $g_{\textrm{qm,s}}/2\pi$=10.0~MHz is obtained. The coupling strength far exceeds the linewidths of the qubit and the magnon, $\gamma_{\textrm{q}}/2\pi = 1.2$~MHz and $\gamma_{\textrm{m}}/2\pi = 1.3$~MHz, respectively. The value is also in a good agreement with the calculated value of 11.8~MHz based on the lowest-order approximation (See \hyperlink{sec:A_static}{Supplementary Information}).

For dynamical control of the magnon quantum state, fast on-off switching of the interaction with the qubit would be useful. In the static coupling scheme, however, it is technically quite challenging to sweep the magnetic field in the time scale faster than the coupling strength. To obtain a tunable coupling between the qubit and the Kittel mode, we adopt a parametrically-induced interaction which has been proposed and demonstrated in superconducting circuits\cite{bib:Wallraff09,bib:Bajjani11} (Fig.~\hyperlink{fig:fig3h}{3a}). In order to suppress the static coupling, a large detuning of 274~MHz between the qubit and the Kittel mode ($\omega_{\textrm{FMR}}/2\pi = 8.410$~GHz) is chosen. Next, we introduce a drive microwave at the mean of the qubit and the Kittel-mode frequencies, i.e., $\omega_{\textrm{d}} = (\omega_{\textrm{FMR}}+\omega_{\textrm{q}})/2$, to induce the parametric coupling. This is basically a third-order nonlinear process; the system absorbs two drive photons and excites the qubit and a magnon simultaneously. The substantial third-order nonlinearity stems from the anharmonicity of the qubit as well as the large coupling strengths $g_{\textrm{q}}$ and $g_{\textrm{m}}$. The interaction Hamiltonian for the parametrically-induced coupling is written as 
\begin{equation}
  {\cal \hat{H}}_{\textrm{qm,p}}/\hbar 
            = g_{\textrm{qm,p}}\,\hat{\sigma}^{+}\,\hat{c}^{\dagger}
            + g_{\textrm{qm,p}}^{*}\,\hat{\sigma}^{-}\,\hat{c}, 
  \label{eq:parametric}
\end{equation}
where the effective coupling strength $g_{\textrm{qm,p}}$ is proportional to the drive microwave power $P_{\textrm{d}}$ [Eq.~(\hyperlink{eq:gqmp}{S7}) in Supplementary Information].

To understand how the continuous-wave spectroscopy reveals the parametrically-induced coupling, we consider the inset of Fig.~\hyperlink{fig:fig3h}{3a} which illustrates the energy levels of the driven hybrid system in the bases of the qubit states $\{ |\textrm{g}\rangle, |\textrm{e}\rangle \}$ and the Kittel-mode magnon-number states $\{ |n_{\textrm{m}}\rangle = |0\rangle, |1\rangle, |2\rangle, \cdots \}$. The two-photon drive induces the Rabi splitting $2g_{\textrm{qm,p}}\sqrt{n_{\textrm{m}}+1}$ between the states $|{\textrm{g}}, n_{\textrm{m}} \rangle$ and $ |{\textrm{e}}, n_{\textrm{m}} + 1 \rangle$. Using a probe microwave with frequency near $\omega_{\textrm{FMR}}$, the change of the magnon excitation spectrum in the driven system is monitored.
The colour intensity plots in Fig.~\hyperlink{fig:fig3h}{3b} show the reflection coefficient as a function of the probe frequency and the parametric drive frequency for several drive powers. When the drive frequency hits the two-photon resonance condition (yellow dashed lines), the dip in the spectrum splits into two. As the drive power increases, an anticrossing feature grows in the spectra, along with an abrupt shift of the magnon frequency (ii $\rightarrow$ iii). The shift from 8.410~GHz to 8.405~GHz is identified as a qubit-state-dependent shift, corresponding to the transition $|{\textrm{g}}, 0\rangle \leftrightarrow |{\textrm{g}}, 1\rangle$  for small $P_{\textrm{d}}$ (i, ii), and $|{\textrm{e}}, 0\rangle \leftrightarrow |{\textrm{e}},1\rangle$ for large $P_{\textrm{d}}$ (iii-vi). This is due to the population transfer from $|\textrm{g},0\rangle$ to $|{\textrm{e}}, 0\rangle$ at large $P_{\textrm{d}}$ as well as the residual static coupling between the qubit and the Kittel mode (See \hyperlink{sec:A_parametric}{Supplementary Information}). The anticrossing observed in the $|{\textrm{e}},0 \rangle \leftrightarrow |{\textrm{e}},1 \rangle $ transition manifests parametrically-induced coupling between the qubit and the magnon. 

As seen in the cross sections presented in Fig.~\hyperlink{fig:fig3h}{3c}, the spacing between the dips, corresponding to $2g_{\textrm{qm,p}}/2\pi$, increases linearly with the drive power. This indicates the capability of arbitral and dynamical coupling between qubit and the Kittel mode via tailoring the parametric drive. The maximum coupling $g_{\textrm{qm,p}}/2\pi$ obtained is 3.4~MHz, which again exceeds the decoherence rates of the qubit and the magnon and assures coherent coupling between them. 

The tunable coupling scheme opens a door to quantum-state engineering of magnon. Starting from the ground state $|{\textrm{g}},0\rangle$, for example, a pulsed parametric drive would generate the single-magnon Fock state (together with a qubit excitation) $|{\textrm{e}},1\rangle$ with a pulse duration $T_{\textrm{p}} = \pi/2g_{\textrm{qm,p}}$. It would also create the qubit-magnon Bell state $(1/\sqrt{2}) \left\{ |{\textrm{g}},0\rangle + |{\textrm{e}},1\rangle \right\} $ with $T_{\textrm{p}} = \pi/4g_{\textrm{qm,p}}$. Such an entangling gate, together with readily applicable single-qubit gates, magnon-state displacement, and qubit readout, enables synthesis and a characterization of arbitrary quantum states of the magnetostatic mode. 

Magnons in a macroscopic-scale ferromagnetic crystal are now ready to be controlled in the quantum regime. This leads us to investigate the ultimate limit of spintronics and magnonics at the single-quantum level. It would also be of particular interest to consider an analogy with the recently advancing optoelectromechanics\cite{bib:Kippenberg13}: Phonons in nanomechanical devices, yet another example of spatially-extended collective excitations in solid, coherently interact both with microwave and optical degrees of freedom, and thus are studied as a candidate for realising quantum transducers between two largely-separated frequency domains\cite{bib:Andrews13,bib:Bochmann13,bib:Bagci14}. Given the demonstrated strong coupling to microwave and the anticipated magneto-optical coupling at infrared, magnons in ferromagnetic insulators may provide an alternative route towards the goal.



%
%
%

\renewcommand{\figurename}{Figure}

\begin{figure*}
  \centering
  \hypertarget{fig:fig1h}{}
  \includegraphics{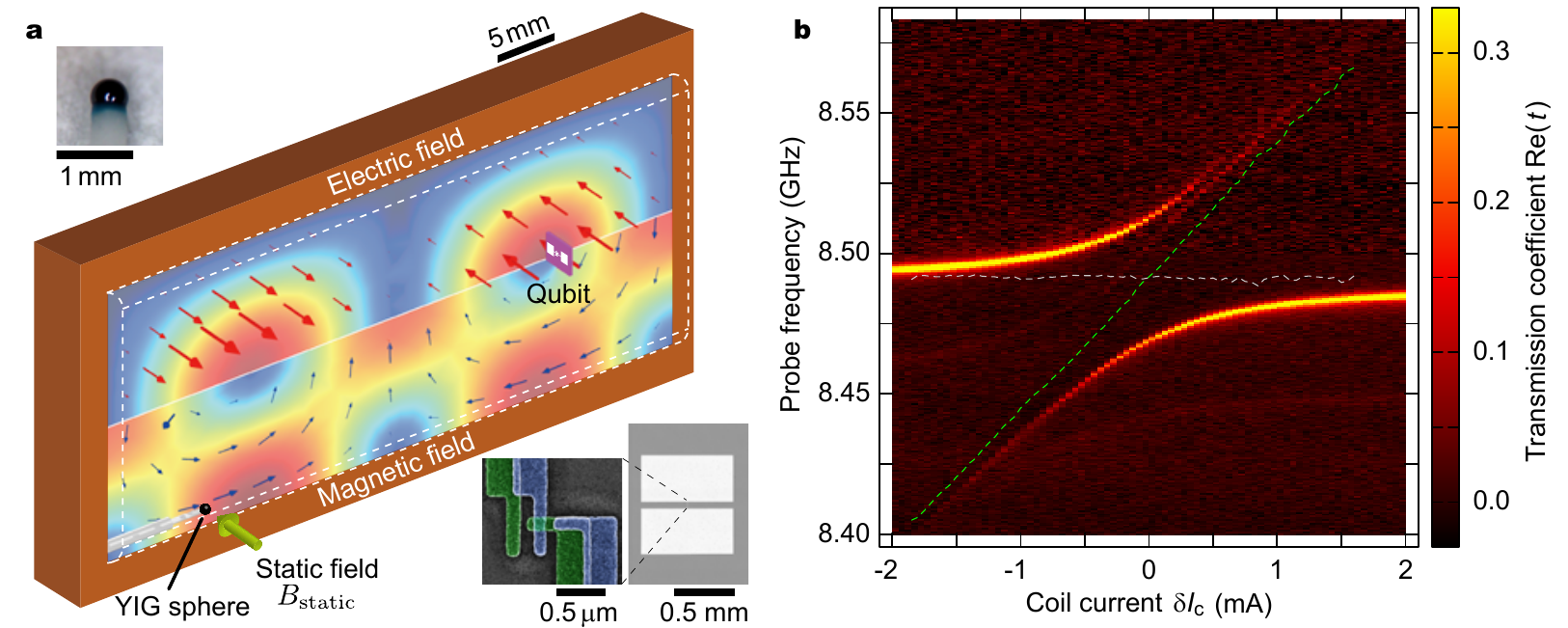}
  \caption{\label{fig:fig1}\textbf{Qubit-magnon hybrid quantum system in a microwave cavity}. \textbf{a}, Simulated microwave field distribution of the TE${}_{102}$ (transverse electric) mode in the cavity. The upper half of the cavity displays the electric field which couples to a transmon-type superconducting qubit (right insets: optical image of the antenna pads and false-colour scanning-electron micrograph of the Josephson junction bridging them). The lower half shows the magnetic field which couples to the spins in a single-crystalline sphere of yttrium iron garnet (YIG; photo in the left inset). A static magnetic field $B_{\textrm{static}}$ around 0.29~T is applied locally to the sphere with a compact magnetic circuit consisting of a pair of permanent magnets and a superconducting coil (Ext.~Fig.~1). %
\textbf{b}, Magnon-photon normal-mode splitting. Amplitude of the microwave transmission coefficient, Re($t$), is measured through the TE${}_{102}$ mode as a function of the probe frequency and the static magnetic field. The field is represented by the relative coil current $\delta I_{\textrm{c}}$ which is defined to be zero at the anticrossing. Here, we use a probe power of $-141$~dBm corresponding to the single-photon average occupancy of the TE${}_{102}$ mode. The anticrossing indicates coherent coupling between the cavity mode and the uniformly precessing magnetostatic mode (Kittel mode) in the sphere. The dashed lines show the TE${}_{102}$-mode (white) and Kittel-mode (green) frequencies obtained from fitting. }
%
\end{figure*}

\begin{figure*}
  \centering
  \hypertarget{fig:fig2h}{}
  \includegraphics{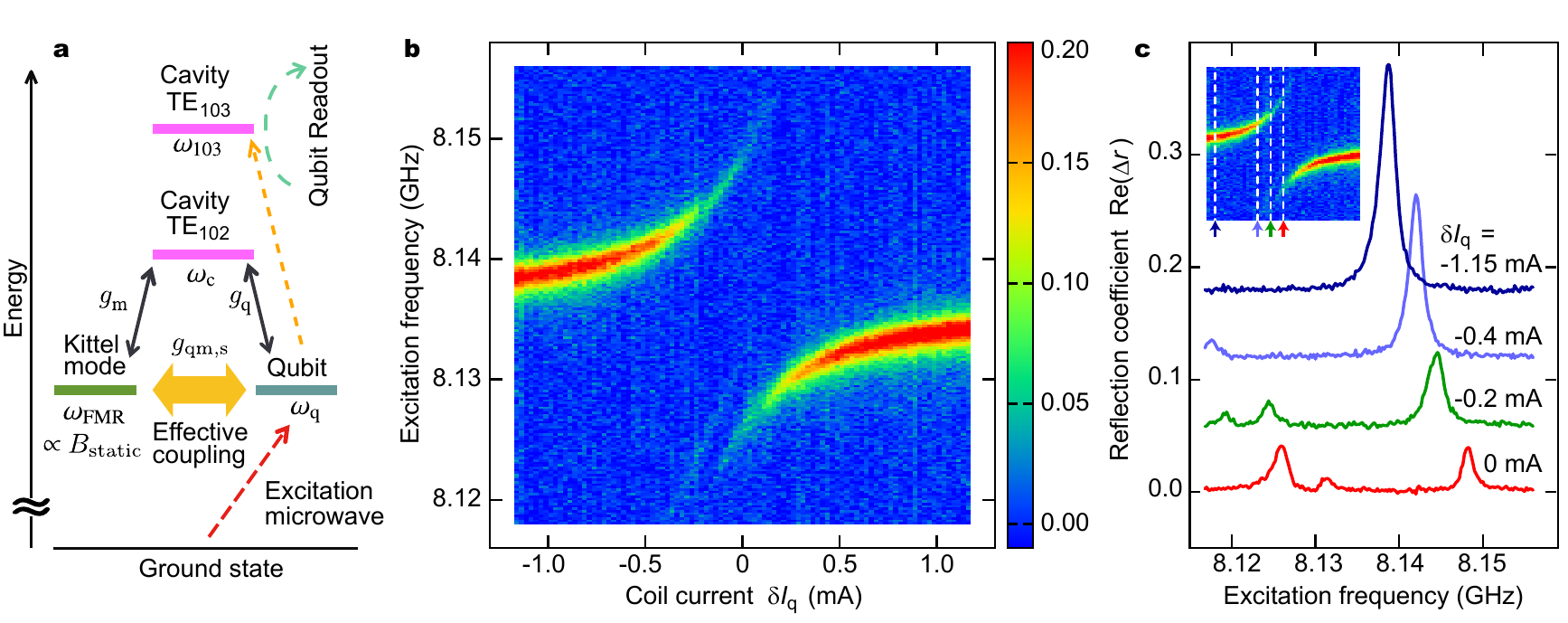}
  \caption{\label{fig:fig2}\textbf{Magnon-vacuum-induced Rabi splitting of the superconducting qubit}. \textbf{a},~Cavity-mediated coupling scheme. The energy diagram illustrates the frequencies, $\omega_{\textrm{c}}$ ($\equiv \omega_{102}$) and $\omega_{103}$, of the cavity TE${}_{102}$ and TE${}_{103}$ modes, the Kittel-mode frequency $\omega_{\textrm{FMR}}$, and the qubit frequency $\omega_{\textrm{q}}$. The qubit couples to the electric field of the TE${}_{102}$ mode with a coupling strength $g_{\textrm{q}}$,  while the Kittel mode in the YIG sphere magnetically couples to the same mode with a coupling strength $g_{\textrm{m}}$. When $\omega_{\textrm{q}} \simeq \omega_{\textrm{FMR}}$, a qubit excitation is transformed to a Kittel-mode magnon and vice versa via the virtual-photon excitation in the TE${}_{102}$ mode. The qubit also couples to the TE${}_{103}$ mode weakly, which results in the dispersive frequency shift of the cavity mode depending on the states of the qubit. The qubit state can be measured through the change in the cavity response to the probe microwave. \textbf{b}, Magnon-vacuum-induced Rabi splitting. Change in the cavity reflection coefficient $\Delta r$ at $\omega_{103}$ is measured  as a function of the qubit excitation frequency and the static magnetic field represented by the relative coil current $\delta I_{\textrm{q}}$. We measure the reflection coefficient with a probe power of $-135$~dBm corresponding to the single-photon average occupancy of the TE${}_{103}$ mode, and with a qubit excitation power of $-$130~dBm corresponding to the qubit Rabi frequency of 1.7~MHz. Another magnetostatic mode weakly couples to the qubit and gives rise to an additional splitting around 8.125~GHz. \textbf{c},~Cross sections of {\textbf{b}} at various static magnetic fields. For clarity, the individual curves are offset vertically by 0.06 each from bottom to top. }
\end{figure*}

\begin{figure*}
  \centering
  \hypertarget{fig:fig3h}{}
  \includegraphics{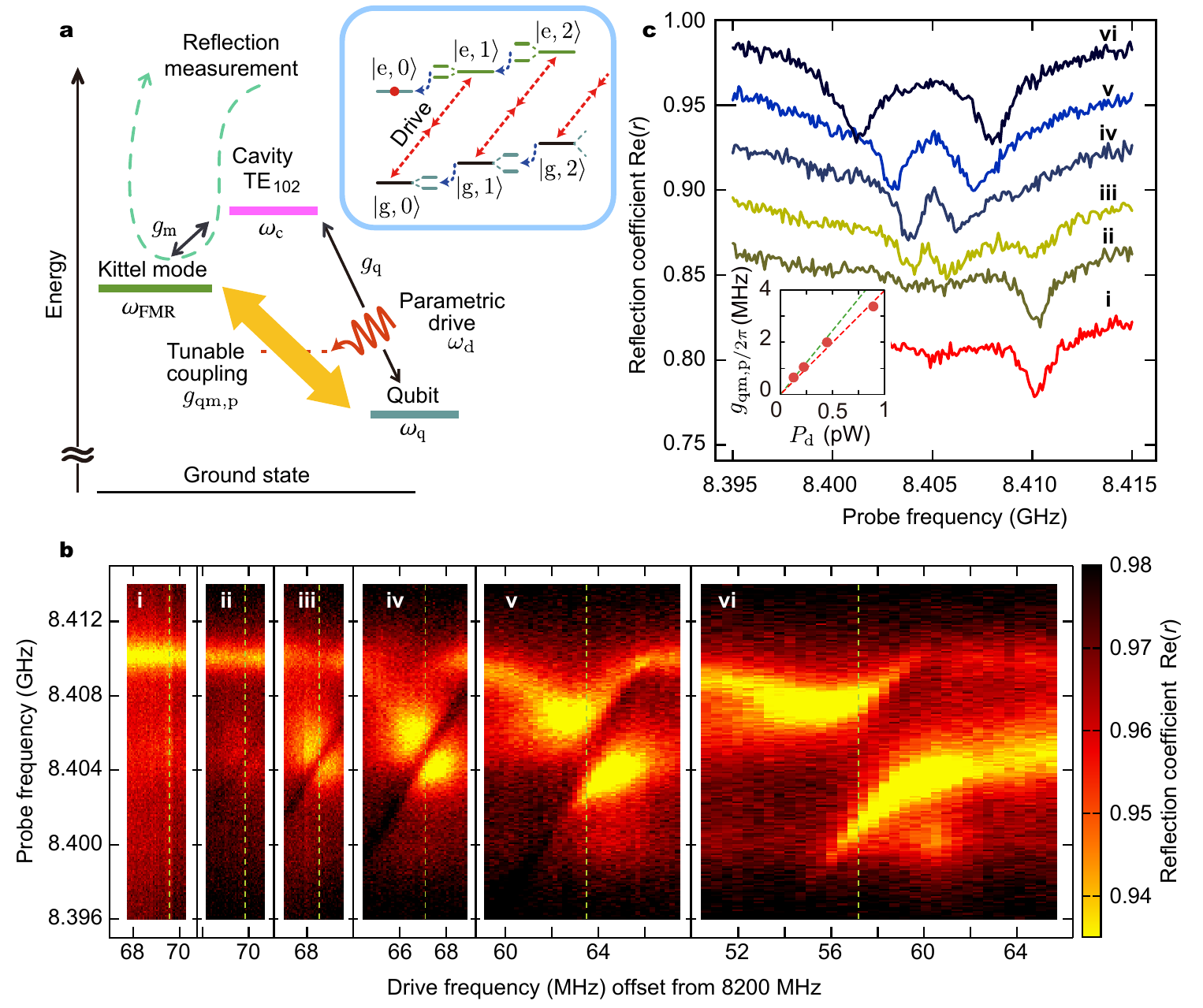}
  \caption{\label{fig:fig3}\textbf{Tunable coupling between superconducting qubit and Kittel mode}. \textbf{a},~Parametrically-induced coupling scheme. The Kittel mode is detuned from the qubit so that the static coupling via the TE${}_{102}$ mode is negligible. An alternative coupling between the qubit and the Kittel mode is parametrically induced by applying a drive microwave at the frequency $\omega_{\textrm{d}} = (\omega_{\textrm{FMR}}+\omega_{\textrm{q}})/2$. The coupling strength $g_{\textrm{qm,p}}$ is controlled by the drive amplitude. The inset shows energy levels labelled with the qubit state ($|\textrm{g}\rangle$ or $|\textrm{e}\rangle$) and the magnon number ($n_\textrm{m} = 0, 1, \cdots$). The parametric drive (red dashed arrows) induces the two-photon transitions, e.g., $|{\textrm {g}},0\rangle \Leftrightarrow |{\textrm{e}},1\rangle$, which results in the Rabi splitting. The blue arrows depict the Kittel-mode decay, and the red circle at $|\textrm{e},0\rangle$ illustrates the dominant steady-state population at higher drive powers. 
\textbf{b},~Kittel-mode spectra indicating tunable coupling with the qubit. The amplitude of the cavity reflection coefficient $\textrm{Re}(r)$ is plotted as a function of the probe and the parametric-drive frequencies. The spectra are measured at the probe power of $-$135~dBm. Small but finite mixing of the Kittel mode with the TE${}_{102}$ mode allows the direct magnon-mode spectroscopy by the probe microwave. Panels i-vi correspond to the drive power $P_{\textrm{d}}$ of $-$104, $-$103, $-$99, $-$96.5, $-$93.5, and $-$90.5 dBm, respectively. \textbf{c},~Cross sections of the plots in {\textbf{b}} at the drive frequencies corresponding to $(\omega_{\textrm{FMR}}+\omega_{\textrm{q}})/2$ (yellow dashed lines in {\bf b}). Note that the frequency depends on the drive power through the ac-Stark shift of the qubit. For clarity, the individual curves are offset vertically by $-0.03$ each from top to bottom. Inset: Coupling strength $g_{\textrm{qm,p}}$ as a function of the drive power $P_{\textrm{d}}$. Also plotted are the linear fit (red dashed line) and the theoretical expectation (green dashed line; See Supplementary Information for the details). 
%
%
}
\end{figure*}

\setcounter{figure}{0}
\renewcommand{\figurename}{Extended Figure}

\begin{figure*}
  \centering
  \hypertarget{fig:exfig1}{}
  \includegraphics{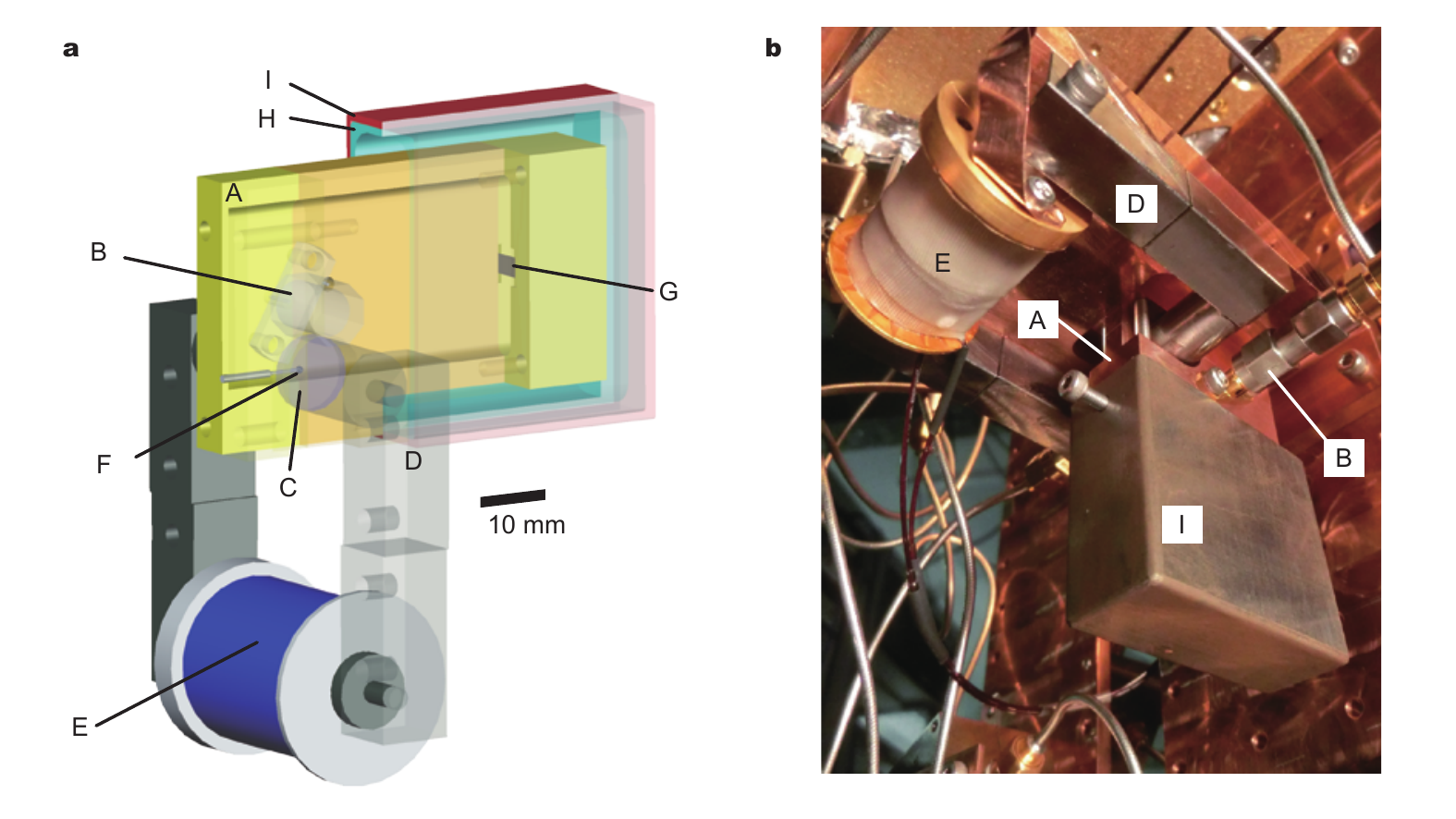}
  \caption{ \textbf{Experimental apparatus}. \textbf{a,} Cut model of the cavity. The elongated rectangular cavity~(A) made of oxygen-free copper has dimensions of 24 $\times$ 3 $\times$ 53~mm. Two SMA connectors~(B; another is on the opposite side) are attached for the microwave transmission and reflection spectroscopy. Couplings of the ports to cavity modes are tuned appropriately by adjusting the length of the connector centre pin protruding into the cavity. A pair of disc-shape neodymium permanent magnets~(C; another is behind the cavity), with a diameter of 10~mm and a thickness of 1.0~mm each, are placed at the ends of a magnetic yoke~(D) made of pure iron. The magnets produce a static field of about 0.29~T at the 4-mm gap in between. The magnetic field can be additionally tuned by the current along a $10^{4}$-turn superconducting coil~(E). The field-to-current conversion ratio is approximately 1.7~T/A. An YIG sphere~(F) glued to an aluminium-oxide rod along the crystal axis $\langle 110 \rangle$ is mounted in the cavity at the centre of the gap between the magnets. The static field is applied in parallel with the crystal axis $\langle 100 \rangle$. A transmon-type superconducting qubit~(G), consisting of two large-area aluminium pads and a single Josephson junction (Al/Al${}_2$O${}_3$/Al), is fabricated on a silicon substrate (See the inset of Fig.~\ref{fig:fig1}a) and is mounted inside the cavity. The qubit and the YIG sphere are separated by 35~mm in the horizontal direction. A double-layer magnetic shield made of aluminium~(H) and pure iron~(I) covers a half of the cavity to protect the qubit from the stray field of the magnet. \textbf{b,} Picture of the apparatus hanging beneath the mixing-chamber plate of a dilution refrigerator. The cavity is cooled down to 10~mK.}
\end{figure*}

\begin{figure*}
  \centering
  \hypertarget{fig:exfig2}{}
  \includegraphics{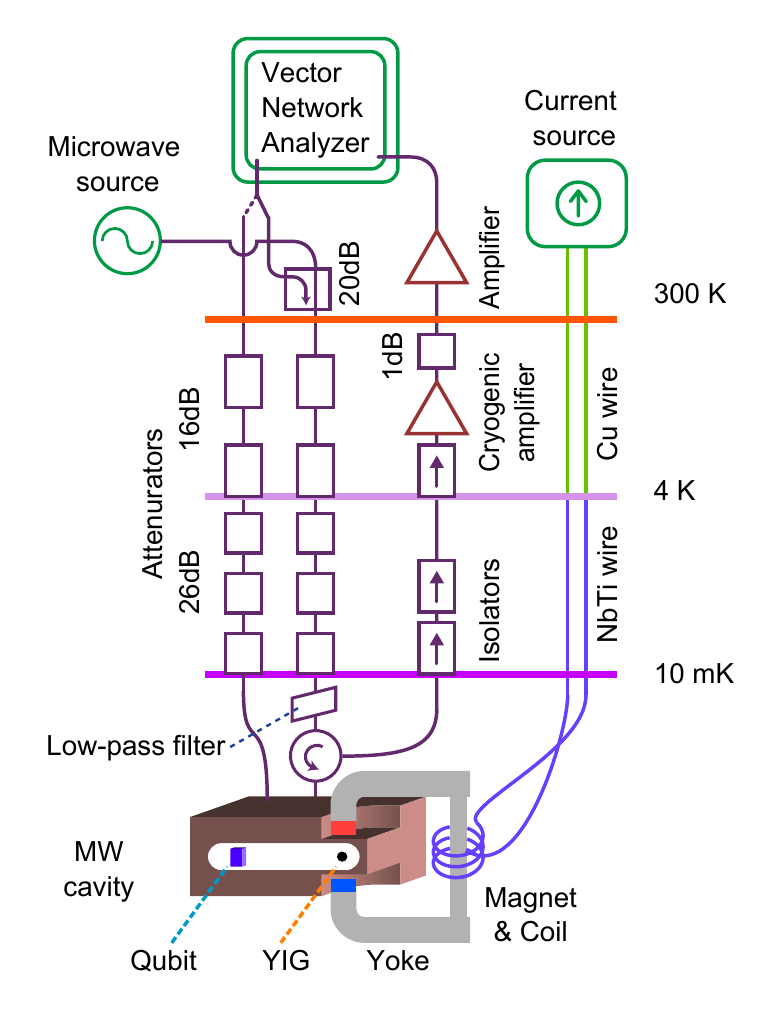}
  \caption{ \textbf{Measurement setup}. The attenuation on the input lines, which at 10~GHz amounts to 49~dB including losses at cables (phosphor-bronze coaxial cables; Coax Corp. SC-119/50-PBC-PBC) and connectors, is enough to prevent the room-temperature thermal noise from reaching the sample space. To further diminish noise above 12~GHz at input port, we introduce a low-pass filter (RLC F-30-12.4-R). For the output line, we place a circulator and isolators (Quinstar XTE0812KCS and XTE0812K), which give the isolation ratio of more than 60~dB. All measurements in this Letter are done with a vector network analyser (VNA; Agilent E5071C). For the static- and parametric-coupling experiments, the excitation microwave generated by a microwave source (Agilent E8247C) is combined with the probe microwave from VNA in a directional coupler (Krytar 120420) and is introduced to the input port. The reflected or transmitted signal from the cavity is amplified in a series of amplifiers at 4~K (Caltech CRYO4-12) and at room temperature (MITEQ AFS4-08001200-09-10P-4). A current source (Yokogawa GS200) is used to tune the static field applied to the YIG sphere. }
\end{figure*}

\begin{figure*}
  \centering
  \hypertarget{fig:exfig3}{}
  \includegraphics{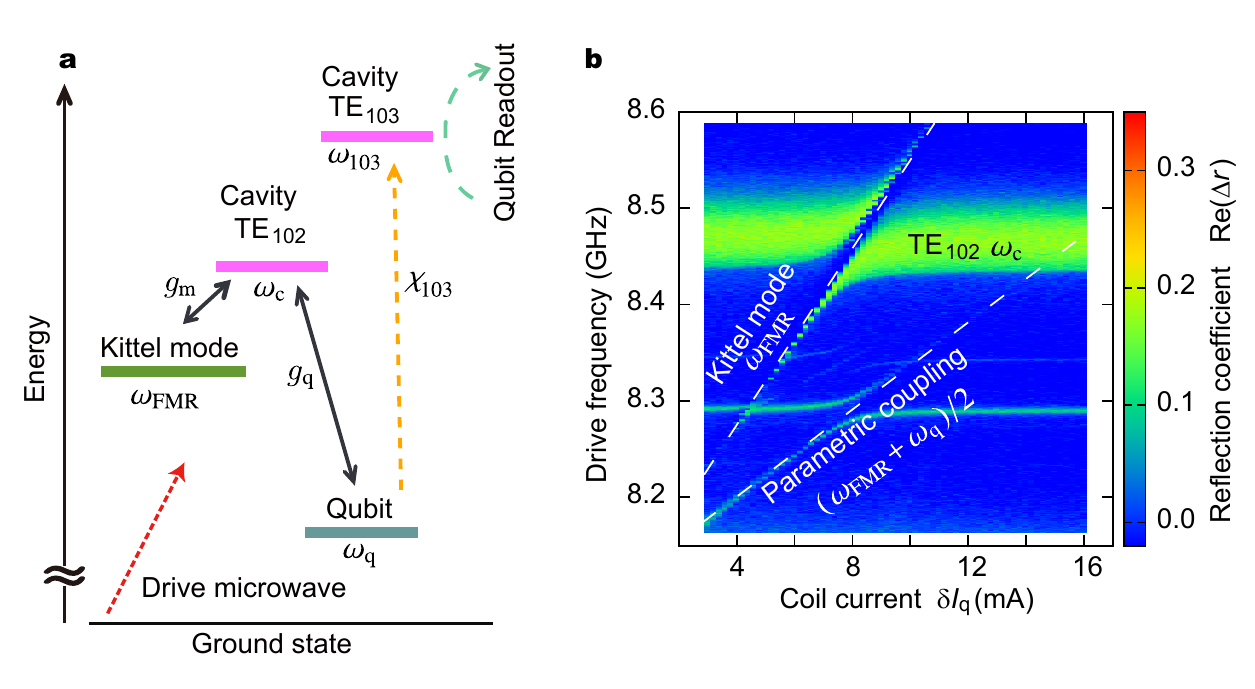}
  \caption{ %
\textbf{Spectroscopy of parametrically-induced transitions}. \textbf{a,} Method of the spectroscopy. The qubit dispersively couples to the cavity TE${}_{103}$ mode, giving rise to the qubit-state-dependent frequency shift $\chi_{103}$. To investigate the energy-level structure of the hybrid system consisting of the qubit, the TE${}_{102}$ mode and the Kittel mode, we search for transitions by driving the system and by simultaneously monitoring the qubit excitation via the frequency shift of the TE${}_{103}$ mode. A few allowed transitions are observed in the low-drive-power limit, while many dipole-forbidden transitions appear when the system is driven strongly. \textbf{b,} Spectrum taken with a strong drive power. Change in the cavity reflection $\Delta r$ at $\omega_{103}$ is measured as a function of the drive frequency and the static magnetic field represented by the coil current $\delta I_{\textrm{q}}$. For a reference, $\delta I_{\textrm{q}}$ is defined to be zero when the Kittel-mode frequency $\omega_{\textrm{FMR}}$ coincides with the qubit frequency $\omega_{\textrm{q}}$. We use the probe and drive microwave power of $-$135~dBm and $-$96.7~dBm, respectively. The line observed at 8.47~GHz, broadened and shifted due to the high-power drive, corresponds to the TE${}_{102}$ mode. Note that the Kittel mode is also detected in the indirect measurement via the qubit because of the residual interaction. The two-photon parametric transition is observed at the frequency $(\omega_{\textrm{FMR}}+\omega_{\textrm{q}})/2$. Other qubit-related multi-photon transitions, e.g., the qubit-TE${}_{102}$-mode parametric transition at $[(\omega_{\textrm{q}}+\omega_{\textrm{c}})/2]/2\pi = 8.29$~GHz, are seen as horizontal lines. %
}
\end{figure*}

\begin{figure*}
  \centering
  \hypertarget{fig:exfig4}{}
  \includegraphics{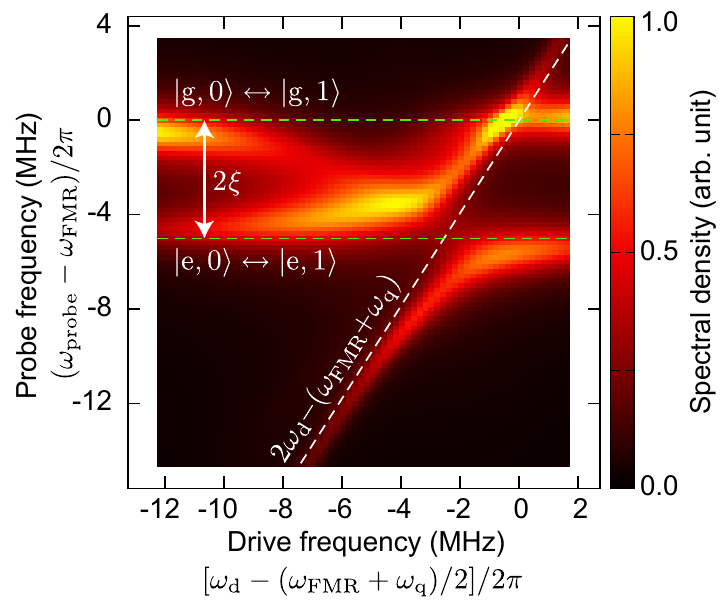}
  \caption{ %
\textbf{Numerical simulation of parametrically-induced coupling}. A simulated Kittel-mode spectrum is plotted as a function of the probe frequency and the parametric-drive frequency. Parameters used for the simulation correspond to those in panel~v of Fig.~\ref{fig:fig3}b (See Supplementary Information). The residual static coupling $\xi$ shifts the Kittel-mode frequency depending on the qubit states ($|\textrm{g}\rangle$ and $|\textrm{e}\rangle$). %
The diagonal dashed line depicts the detuning of the parametric drive. Away from the two-photon parametric resonance $|\textrm{g},0\rangle \Leftrightarrow |\textrm{e},1\rangle$, only the Kittel-mode excitation $|\textrm{g},0\rangle \leftrightarrow |\textrm{g},1\rangle$ (upper horizontal dashed line) appears since the qubit is in the ground state. As the drive frequency approaches the resonance, parametric drive excites the qubit together with a Kittel-mode magnon, leading to a shift of the Kittel-mode spectrum to the line $|\textrm{e,0}\rangle \leftrightarrow |\textrm{e},1\rangle$ (lower horizontal dashed line). The dressed states between $|\textrm{g},0\rangle$ and $|\textrm{e},1\rangle$ give rise to an avoided crossing at the resonance, indicating the parametrically-induced coherent coupling (Inset of Fig.~\ref{fig:fig3}).
}
\end{figure*}

\clearpage
\renewcommand{\theequation}{S\arabic{equation}}

\onecolumngrid

\begin{quotation}
\begin{center}
{\large \textbf{Supplementary Information for ``Coherent coupling between ferromagnetic magnon and superconducting qubit''}} \newline \newline
Y. Tabuchi${}^{1}$, S. Ishino${}^{1}$, A. Noguchi${}^{1}$, T. Ishikawa${}^{1}$, R. Yamazaki${}^{1}$, K. Usami${}^{1}$, and Y. Nakamura${}^{1,2}$ \newline
{\small \textit{${}^{1}$Research Center for Advanced Science and Technology (RCAST),}} \\
{\small \textit{The University of Tokyo, Meguro-ku, Tokyo 153-8904, Japan \ \ and}} \\
{\small \textit{${}^{2}$Center for Emergnent Matter Science (CEMS), RIKEN, Wako, Saitama 351-0198, Japan}}
\end{center}
\end{quotation}

\twocolumngrid
\section*{Experimental apparartus}
An illustrative drawing and a photo of the experimental apparatus are shown in Ext.~Fig.~\hyperlink{fig:exfig1}{1}. The wiring, circuit components and instruments used in the experiments are shown in Ext.~Fig.~\hyperlink{fig:exfig2}{2}. Each of the microwave powers presented in the paper refers to the one at the relevant cavity port. 

\hypertarget{sec:A_qubit_and_cavity}{}
\section*{Qubit and cavity modes}
We use a transmon-type qubit with the bare qubit frequency $\omega_{\textrm{q,bare}}/2\pi$ and the anharmonicity $\alpha/2\pi$ of 8.204~GHz and $-158$~MHz, respectively. The anharmonicity $\alpha$ is defined as the difference $\omega_{\textrm{ef}}-\omega_{\textrm{ge}}$ between the bare transition frequencies of the ground state to the first excited state $\omega_{\textrm{ge}}$ ($\equiv \omega_{\textrm{q,bare}}$) and the first excited state to the second excited state of the qubit $\omega_{\textrm{ef}}$. The qubit couples to the cavity modes TE${}_{101}$, TE${}_{102}$ and TE${}_{103}$, with coupling strengths of 85~MHz, 121~MHz and 142~MHz, respectively. Owing to the coupling to the cavity field, the qubit frequency $\omega_{\textrm{q}}/2\pi$ is shifted by $-$71~MHz (Lamb shift) from $\omega_{\textrm{q,bare}}/2\pi$. 

The cavity has the bare mode frequencies ($\omega_{101,\textrm{bare}}/2\pi$, $\omega_{102,\textrm{bare}}/2\pi$ and $\omega_{103,\textrm{bare}}/2\pi$) of 6.993~GHz, 8.420~GHz and 10.452~GHz. They are subject to a shift due to coupling with the qubit by $-6$~MHz, $+68$~MHz, $+9$~MHz, respectively. We tune the coupling of the cavity modes to the input and output ports by adjusting the length of the centre pin protruding into the cavity. The input and output couplings for the modes TE${}_{102}$ and TE${}_{103}$ are $\kappa_{\textrm{in},102}/2\pi = 0.68$~MHz, $\kappa_{\textrm{out},102}/2\pi = 0.53$~MHz, $\kappa_{\textrm{in},103}/2\pi = 0.15$~MHz and $\kappa_{\textrm{out},103}/2\pi = 2.72$~MHz. The internal loss of the modes TE${}_{102}$ and TE${}_{103}$ are 1.26~MHz and 1.24~MHz, respectively. 

Throughout the paper, we use the renormalized frequencies for the qubit $\omega_{\textrm{q}}$ and the cavity modes ($\omega_{\text{101}}$,  $\omega_{\text{102}} \equiv \omega_{\text{c}}$, and $\omega_{103}$), taking into account the frequency shifts due to the interaction. Namely, $\omega_{\textrm{q}}/2\pi$ = 8.136~GHz, $\omega_{\textrm{101}}/2\pi$ = 6.987~GHz, $\omega_{102}/2\pi$ ($\equiv \omega_{\textrm{c}}/2\pi$) = 8.488~GHz and $\omega_{103}/2\pi$ = 10.461~GHz.

\section*{YIG sample}

We purchased the YIG single crystal from Ferrisphere Inc.~\cite{bib:Ferrisphere} The spherical sample is polished to the surface roughness below 50~nm and glued to an aluminium-oxide rod at the factory. We apply the static magnetic field $B_{\textrm{static}}$ to the $\langle 100\rangle$ crystal axis. 
\\

\hypertarget{sec:A_static}{}
\section*{Qubit-magnon coupling strength}
The effective coupling between the qubit and the Kittel mode $g_{\textrm{qm,s}} = g_{\textrm{q}}g_{\textrm{m}}/\Delta$ should be calculated using the bare qubit $\omega_{\text{q,bare}}$ and cavity $\omega_{\textrm{102,bare}}$ frequencies to avoid double-counting of the effect of interactions in the perturbative treatment. Therefore, we use the detuning $\Delta/2\pi = (\omega_{\textrm{q,bare}}-\omega_{102,\textrm{bare}})/2\pi$ of 215~MHz, and evaluate the coupling to be $g_{\textrm{qm,s}}/2\pi = 11.8$~MHz. 

\hypertarget{sec:A_parametric}{}
\section*{Parametrically induced coupling}

\paragraph{Spectroscopy} To find parametrically-induced transitions, we perform spectroscopy of the hybrid system. Extended Figure~\hyperlink{fig:exfig3}{3a} illustrates the method. We use the TE${}_{103}$ mode for the qubit readout; when the drive microwave hits transitions related to the qubit, the cavity reflection coeffcient at $\omega_{103}$ changes. 

In Ext.~Fig.~\hyperlink{fig:exfig3}{3b}, the upper diagonal dashed line corresponds to the Kittel-mode frequency, which linearly moves along with the coil current and splits when the frequency coincides with the TE${}_{102}$-mode frequency. (Note that the Kittel mode is also observed by the direct measurement through the TE${}_{102}$ mode as shown in Fig.~\hyperlink{fig:fig1h}{1b}.) On the other hand, the lower dashed line is ascribed to the parametrically-induced transition. The transition frequency is equal to $(\omega_{\textrm{q}}+\omega_{\textrm{FMR}})/2$. The transition frequency and the Kittel mode frequency $\omega_{\textrm{FMR}}$ meet each other at the qubit frequency $\omega_{\textrm{q}}$. 

\paragraph{Coupling strength}

The parametrically-induced coupling strength $g_{\textrm{qm,p}}$ is derived from a Hamiltonian taking into account the TE${}_{102}$ mode, the qubit and the Kittel mode, which is expressed as
\begin{eqnarray}
  {\cal \hat{H}}/\hbar &=& 
     \omega_{102,\textrm{bare}}   \hat{a}^\dagger \hat{a} 
   + \omega_{\textrm{FMR,bare}} \hat{c}^\dagger \hat{c} 
     \nonumber \\
  &+&
    \left(
      \omega_{\textrm{q,bare}}-\frac{\alpha}{2}
    \right) \hat{b}^\dagger \hat{b}
   + \frac{\alpha}{2}\left(\hat{b}^\dagger \hat{b}\right)^2 \nonumber \\
&+& g_{\textrm{q}}(\hat{a}^\dagger \hat{b} + \hat{a} \hat{b}^\dagger) 
+ g_{\textrm{m}}(\hat{a}^\dagger \hat{c} + \hat{a} \hat{c}^\dagger) \nonumber \\
&+& \sqrt{\kappa_{102} \frac{P_{\textrm{d}}}{\hbar\omega_{\textrm{d}}}}
\left[
 \hat{a}^\dagger \exp\left(-i\omega_{\textrm{d}}\right)
+\hat{a} \exp\left(i\omega_{\textrm{d}}\right)\right],
\end{eqnarray}
where $\hat{a}$ and $\hat{b}$ are the annihilation operators of the photon in the TE${}_{102}$ mode and a transmon qubit, $\alpha$ is the anharmonicity of the qubit, and $\kappa_{102} = \kappa_{\textrm{out},102}$ since we introduce the microwave drive through the cavity output port. Here, we define the lowest two levels of the anharmonic oscillator as the qubit subspace. Provided we explicitly write the lowering operator as $\hat{b} = \sum_{n \geq 0} \sqrt{n+1}|n\rangle \langle n+1|$, the qubit lowering operator is defined as $\hat{\sigma}^{-} = (\hat{\sigma}^{+})^\dagger= |0\rangle \langle 1|$. 
By applying successive unitary transformations to diagonalise the Hamiltonian\cite{bib:aveHamil} and using approximation in which higher order terms are safely truncated, we obtain the effective Hamiltonian
\begin{eqnarray}
  {\cal \hat{H}}_{\textrm{eff}}
    &=& {\cal \hat{H}}_{\textrm{qm,p}} + {\cal \hat{H}}_{\textrm{res}}, \label{eq:Hamiltot}
\end{eqnarray}
where ${\cal \hat{H}}_{\textrm{qm,p}}$ and ${\cal \hat{H}}_{\textrm{res}}$ represent a parametrically-induced tunable coupling and a residual static coupling between the qubit and the Kittel mode, respectively. They are written as
\begin{eqnarray}
  {\cal \hat{H}}_{\textrm{qm,p}}/\hbar
    &=& 
 g_{\textrm{qm,p}}\,e^{i(\omega_{\textrm{q}}+\omega_{\textrm{FMR}}+2\xi-2\omega_{\textrm{d}})t}\,\hat{\sigma}^{+}\,\hat{c}^{\dagger} \nonumber \\
    &+& 
 g_{\textrm{qm,p}}^{*}\,e^{-i(\omega_{\textrm{q}}+\omega_{\textrm{FMR}}+2\xi-2\omega_{\textrm{d}})t}\,\hat{\sigma}^{-}\,\hat{c} \label{eq:Hamiltc} \\
  {\cal \hat{H}}_{\textrm{res}}/\hbar
     &=& 2 \xi\,\hat{c}^\dagger \hat{c} \hat{\sigma}^{+}\hat{\sigma}^{-}.
       \label{eq:Hamilres}
\end{eqnarray}
The coupling strength $g_{\textrm{qm,p}}$ is expressed as \hypertarget{eq:gqmp}{}
\begin{eqnarray}
  g_{\textrm{qm,p}} &\simeq& 
    2 g_{\textrm{q}} g_{\textrm{m}}^{*} 
   \left(
        \frac{\kappa_{102}P_{\textrm{d}}}{\hbar\omega_{\textrm{d}}}
   \right)
   \left(
      \frac{g_{\textrm{q}}}{\Delta_{\textrm{d}}}
   \right)^2 \nonumber \\
  &\times& 
   \Biggl[ 
    \frac{5/6}{(\Delta_{\textrm{q}}-\Delta_{\textrm{m}})^2 \Delta_{\textrm{m}}} 
   \nonumber \\
  &-& 
    \frac{1}{(\Delta_{\textrm{q}}-\Delta_{\textrm{m}})
             (\Delta_{\textrm{q}}-\Delta_{\textrm{m}}-\alpha) \Delta_{\textrm{m}}}
    \nonumber \\
  &+&
    \frac{2/3}{(\Delta_{\textrm{q}}-\Delta_{\textrm{m}})
               (\Delta_{\textrm{q}}+\Delta_{\textrm{m}}) 
               \Delta_{\textrm{m}}}
    \nonumber \\
  &+&
    \frac{4/3}{(\Delta_{\textrm{q}}-\Delta_{\textrm{m}})^2 
              (\Delta_{\textrm{q}}+\Delta_{\textrm{m}})}
    \nonumber \\
  &+&
    \frac{1/3}{\Delta_{\textrm{q}}
              (\Delta_{\textrm{q}}+\Delta_{\textrm{m}})
               \Delta_{\textrm{m}}}
    \nonumber \\
  &+&
    \frac{1/2}{(\Delta_{\textrm{q}}-\Delta_{\textrm{m}}-\alpha)
              (\Delta_{\textrm{q}}-\Delta_{\textrm{m}}-2\alpha)
               \Delta_{\textrm{m}}}
    \nonumber \\
  &+&
    \frac{1/6}{\Delta_{\textrm{q}}
              (\Delta_{\textrm{q}}-\Delta_{\textrm{m}})
               \Delta_{\textrm{m}}}
  +
    \frac{1/2}{\Delta_{\textrm{q}}
              (\Delta_{\textrm{q}}-\Delta_{\textrm{m}})^2}
    \nonumber \\
  &+&
    \frac{2/3}{\Delta_{\textrm{q}}
              (\Delta_{\textrm{q}}-\Delta_{\textrm{m}})
              (\Delta_{\textrm{q}}+\Delta_{\textrm{m}})}
    \nonumber \\
  &-&
    \frac{1/6}{(\Delta_{\textrm{q}}-\Delta_{\textrm{m}})
               (\Delta_{\textrm{q}}-\Delta_{\textrm{m}}-2\alpha)
                \Delta_{\textrm{m}}}
    \nonumber \\
  &-&
    \frac{1/2}{(\Delta_{\textrm{q}}-\alpha)
               (\Delta_{\textrm{q}}-\Delta_{\textrm{m}})
               (\Delta_{\textrm{q}}-\Delta_{\textrm{m}}-\alpha)}
    \nonumber \\
  &-&
    \frac{1/12}{(\Delta_{\textrm{q}}-\alpha)
                (\Delta_{\textrm{q}}-\Delta_{\textrm{m}})
                 \Delta_{\textrm{m}}}
    \nonumber \\
  &+&
    \frac{1/12}{(\Delta_{\textrm{q}}-\alpha)
                (\Delta_{\textrm{q}}-\Delta_{\textrm{m}}-2\alpha)
                 \Delta_{\textrm{m}}}
    \nonumber \\
  &-&
    \frac{1/6}{(\Delta_{\textrm{q}}-\alpha)
                (\Delta_{\textrm{q}}-\Delta_{\textrm{m}})
                (\Delta_{\textrm{q}}-\Delta_{\textrm{m}}-2\alpha)}
    \Biggr], 
\end{eqnarray}
and the residual coupling strength $\xi$ is denoted as 
\begin{eqnarray}
  \xi 
  &\simeq& -|g_{\textrm{q}}|^2 |g_{\textrm{m}}|^2 
    \nonumber \\
  &\times& \Biggl [ 
    \frac{5/12}{ \Delta_{\textrm{q}}
                (\Delta_{\textrm{q}}-\Delta_{\textrm{m}})
                 \Delta_{\textrm{m}}}
  +
    \frac{5/12}{(\Delta_{\textrm{q}}-\Delta_{\textrm{m}})
                 \Delta_{\textrm{m}}^2}
    \nonumber \\
  &+& 
    \frac{1/4}{\Delta_{\textrm{q}} \Delta_{\textrm{m}}^2}
  +
    \frac{1/12}{\Delta_{\textrm{q}}^2 \Delta_{\textrm{m}}}
  +
    \frac{1/12}{ \Delta_{\textrm{q}}^2
                (\Delta_{\textrm{q}}-\Delta_{\textrm{m}})}
    \nonumber \\
  &+&
    \frac{5/12}{(\Delta_{\textrm{q}}-\Delta_{\textrm{m}})^2
                 \Delta_{\textrm{m}}}
  -
    \frac{1/12}{(\Delta_{\textrm{q}-\alpha})
                 \Delta_{\textrm{m}}^2}
    \nonumber \\
  &-&
    \frac{5/12}{(\Delta_{\textrm{q}}-\alpha)
                (\Delta_{\textrm{q}}-\Delta_{\textrm{m}}-\alpha)
                 \Delta_{\textrm{m}}}
    \nonumber \\
  &-&
    \frac{5/12}{(\Delta_{\textrm{q}}-\Delta_{\textrm{m}}-\alpha)
                 \Delta_{\textrm{m}}^2}
  - 
    \frac{1/12}{(\Delta_{\textrm{q}}-\alpha)^2 
                 \Delta_{\textrm{m}}}
    \nonumber \\
  &-&
    \frac{1/12}{(\Delta_{\textrm{q}}-\alpha)^2
                (\Delta_{\textrm{q}}-\Delta_{\textrm{m}}-\alpha)}
    \nonumber \\
  &-&
    \frac{5/12}{(\Delta_{\textrm{q}}-\Delta_{\textrm{m}}-\alpha)^2
                 \Delta_{\textrm{m}}}
    \Biggr].
    \label{eq:strength_res}
\end{eqnarray}
Here, we define 
$\Delta_{\textrm{q}} = \omega_{102,\textrm{bare}}-\omega_{\textrm{q,bare}}+%
2g_{\textrm{q}}/(\omega_{102,\textrm{bare}}-\omega_{\textrm{q,bare}})$, 
$\Delta_{\textrm{m}} = \omega_{102,\textrm{bare}}-\omega_{\textrm{FMR}}+%
g_{\textrm{q}}/(\omega_{102,\textrm{bare}}-\omega_{\textrm{q,bare}})$, and
$\Delta_{\textrm{d}} = \omega_{102,\textrm{bare}}-\omega_{\textrm{d}}+%
g_{\textrm{q}}/(\omega_{102,\textrm{bare}}-\omega_{\textrm{q,bare}})$. 
The conversion ratio $g_{\textrm{qm,p}}/P_{\textrm{d}}$ between the drive power and the coupling strength in our experiment is estimated to be 4.9~MHz/pW, which is indicated in the inset of Fig.~\hyperlink{fig:fig3h}{3c} as a green dashed line. 
The discrepancy between the theoretical~(4.9~MHz/pW) and the experimental (4.0$\pm$0.4~MHz/pW) values is attributed to higher-order terms dropped in the secular approximation. 

\paragraph{Residual coupling}

In the tunable-coupling experiment, we suppress the static coupling between the qubit and the Kittel mode, $g_{\textrm{qm,s}}$, by setting a relatively large detuning of 274~MHz. However, the residual coupling ${\cal \hat{H}}_{\textrm{res}}$ remains finite, which shifts the Kittel-mode frequency depending on the qubit state ($|\textrm{g}\rangle$ or $|\textrm{e}\rangle$). We obtain the residual coupling strength $\xi/2\pi$ of $-$2.5~MHz from the observation of the Kittel-mode doublet in the presence of qubit excitation, whereas the calculated value from Eq.~(\ref{eq:strength_res}) is $-$1.2~MHz. 

In order to understand the effect of the residual coupling, we perform a numerical simulation of the Kittel-mode spectrum under the parametric drive. Parameters are chosen such that they reproduce the result shown in panel v of Fig.~\hyperlink{fig:fig3h}{3b}: the residual coupling $\xi/2\pi = -2.5$~MHz, the coupling strength $g_{\textrm{qm,p}}/2\pi$ = 2.2~MHz, the Kittel mode linewidth $\gamma_{\textrm{m}}/2\pi$ = 1.3~MHz, the qubit relaxation rate $\gamma_{\textrm{q}}/2\pi$ = 0.13~MHz and the qubit dephasing rate $\gamma_{\phi}/2\pi$ = 0.3~MHz. 

Extended Figure~\hyperlink{fig:exfig4}{4} shows the simulated Kittel-mode spectrum as a function of the probe frequency and the drive frequency. The simulation reveals that the residual coupling ${\cal \hat{H}}_{\textrm{res}}$ shifts the Kittel-mode frequency depending on the qubit state. As the parametric drive excites the qubit and the Kittel mode simultaneously, the Kittel-mode frequency is shifted by $2\xi$. At the resonance it further shows two-photon Rabi splitting. The simulation reproduces the characteristic features in Fig.~\hyperlink{fig:fig3h}{3b} pretty well.

\end{document}